\documentclass[10pt,journal,compsoc]{IEEEtran}

\ifCLASSOPTIONcompsoc
  \usepackage[nocompress]{cite}
\else
  \usepackage{cite}
\fi
\usepackage{hyperref}
\usepackage{graphicx}

\ifCLASSINFOpdf
\else
\fi
\usepackage{color}
\usepackage{listings}
\begin{document}
%
\title{The Little Phone That Could Ch-Ch-Chroot}
%
%
%
%
\author{Jack~Whitter-Jones
        and~Mathew~Evans
\IEEEcompsocitemizethanks{\IEEEcompsocthanksitem Jack Whitter-Jones is a PhD candidate within the Faculty
of Computing, Engineering and Mathematics at the University of South Wales\protect\\
E-mail: jack.whitter-jones@southwales.ac.uk
\IEEEcompsocthanksitem Mathew Evans is a Researcher within the field of Computer Forensics.\protect\\
E-mail: munk@nop.ninja
}
\thanks{}}
%
%

\markboth{BSides London, 2019}%
{Shell \MakeLowercase{\textit{et al.}}: The Little Phone That Could Ch-Ch-Chroot.}
%



\IEEEtitleabstractindextext{%
\begin{abstract}
Security testing has been a career path that many are beginning to take. In doing so, security testing can hit the realms of many different types of engagements, ranging from web, infrastructure and social engineering. With the risk of sabotage, corporate espionage it has been seen that many organisations are beginning to develop a tactical capability. In doing so, the term 'Red Team' has been coined to market such engagements. Red Teaming is the method of having almost free reign towards a target. By doing such an engagement a target will be able to fully understand the breadth of vulnerabilities facing their organisation. However, Red Teaming can be an expensive and resource intensive task. Through this paper, it is discussed that it is possible to make a covert disposable phone to help aide Red Teamer's with the reconnaissance phase without drawing attention to themselves within a day to day task.

\end{abstract}

}

\maketitle

\IEEEdisplaynontitleabstractindextext

%
\IEEEpeerreviewmaketitle

\ifCLASSOPTIONcompsoc
\IEEEraisesectionheading{\section{Introduction}\label{sec:introduction}}
\else
\section{Introduction}
\label{sec:introduction}
\fi

%
%
%
%
\IEEEPARstart{S}{ecurity} engagements, hacking, penetration testing all encompass the overall posture in how to exploit an organisations digital system to achieve mission success, may it be providing an overall security posture or something more sinister. There is a range of different types of security engagements can encompass many different areas of security. An assessment can encompass, Web Application, Red Teaming, Infrastructure and Social Engineering Testing. Any one of these engagement types can cost an extremely large amount of money, both for the provider and the client. Of interest to this paper is the Red Team Engagement process. This paper aims to help develop the area of Red Teaming's, physical processes through the development of a covert phone. The paper is split first for the reasons why we chose to do build the device, and then the key implementation of the device and then finally concluding the paper with a discussion.

\subsection{What is Hacking and Red Teaming?}
"Hacking" is the term used to describe as, "to circumvent security and break into (a network, computer, file, etc.), usually with malicious intent"\cite{Hacking}. This however, can also be associated to social engineering. Social engineering is the art of exploiting a person or persons to gather intelligence for a mission’s objective. Red Teaming however, is the process in which the overall network and staff are tested to better understand, identifying and assessing vulnerabilities that face an organisation. Red Teaming can provide an extremely detailed understanding of the attack surface. This is due to depth and breadth of a test as they are tailored to a scenario towards the customers’ requirements.

\section{Device Build}
\label{sec:phone}
Within this project, the requirement was to distinguish what was 'expensive' in which a device for less than \textsterling35.00 was a goal. Additionally, the device would be required to be able to run the overall architecture specific requirements had to be met. The following requirements were necessary to be able to allow for the end implementation:
\begin{itemize}
    \item Android - Openness for underlying architectures.
    \item Cheap - Ability to purchase a device for plug and play features.
    \item Rootable - The underlying concepts require a Rootable architecture.
    \item Deniable - The device must provide a level of deniability to allow the user not to be tied to an engagement.
    \item Ease of Use - The ability to change, update and maintain quickly.
\end{itemize}
Meeting these five requirements allows for a versatile and almost limitless set of potential outcomes. In addition, the baseline that it provides can be applied to other Android based devices with varying ranges of functionality e.g., Smart Watches. As stated the phones underlying architecture utilises the Android stack. It is worth noting that this implementation used was Android Version 6.0.1, Kernel 3.18.19. The base handset was an Alcatel, PIXI Volcano Black R-B, 4034X, which cost \textsterling29.99 during the latter half of 2017. The Alcatel met the full requirements above. A comparable setup to the Alcatel phone would be a 3G mobile module, USB dongle utilised by a Raspberry PI 3. This setup would cost the same if not more, and would look suspicious when carried around.

\section{Implementation}
\label{sec:implementation}
This section outlines the thought processes and objectives that were required to be able to implement an architecture to achieve the projects mission statement, which is a cheap disposable device, that has an implementation for a high level of plausible deniability and limiting in evidential integrity.

\subsection{SELinux}
SELinux is prevalent across many Android devices it is worth providing a high-level background understanding to it. SELinux is a mandatory access control facility that provides the overall security architecture for Android. This is on top of Android's sandbox and discretionary access control (DAC). SELinux is required to be utilised almost exclusively across the phones life cycle, thus removing SELinux isn't allowed. This is partly down to the requirements of removing the SELinux implementation within the firmware of the device, with the added problem of discrediting deniability towards the device operator. SELinux implements two approaches, enforcement or passive \cite{RedhatSeLinux}.
\begin{itemize}
    \item Enforcing - SELinux policy is enforced. SELinux denies access based on SELinux policy rules.
    \item Permissive - SELinux policy is not enforced. SELinux does not deny access, but denials are logged for actions that would have been denied if running in enforcing mode. 
\end{itemize}
To provide further granularity, SELinux provides contextual security to processes. This is done through security contexts which can be applied to users and objects, objects are typically referred to as files. As these files situate within the Android operating system, persistence is associated to such files. A objects security context is immutable throughout the Android architecture. As objects are somewhat immutable, the removal of SELinux isn't fully possible as the Android operating system will re-implement SELinux if the policy or context is subverted.

\subsection{Device Rooting}
Rooting the device was straight forward as much of the work had been done and written up on the XNA website, which can be found at \cite{XNA}. Once rooted, the device was loaded with CyanogenMod version 13. CyanogenMod 13 provided a sustainable base for the base architecture to be developed on. CyanogenMod is commonly utilised as a mobile enthusiast’s base for many devices during the time of creating this project. This provides a suitable platform to build our plausible deniability as anyone carrying such a device would be considered an advocate for technology.

\subsection{CPIO \& Chroot}
\label{CPIO}
The concept of using a Chroot within an Android is far from new. There are examples such as NetHunter the Kali install that uses Chroot \cite{Kali}, and Debian is commonly installed via a Chroot into Android phones. Additionally, the use of CPIO archive files to store Linux firmwares is also not new, it is a common practice used on routers and small embedded devices. 

Chroot give the ability to have a different sub-system running almost independently. Allowing features not present in a system to be made available. The choice to use Tinycorelinux presented a few advantages, over previous examples that have used Debian or other Linux distributions. The first major advantage is a drastic size reduction. A minimum Debian install requires 350mb of space, a typical minimum install of NetHunter takes 1gb of space, TinyCoreLinux's minimum install can be as little as 32mb \cite{TinyCore}.

By placing the full distribution into a CPIO format natively supported by Android, but unsupported by most mobile forensics packages, it is possible to get a first layer of protection. CPIO also applies GZ Compression which protects against data carvers. Finally, to hide the CPIO without the embedded system being easily found as just another file. It is stored at the end of the user data partition using the "dd" tool, which is also included in Android. 

Finally, with SELinux disabled the onboot, the userinit.sh script can create a memory only temporary filesystem, extract the Linux sub-system from the hidden CPIO, and activate the Chroot. 
\begin{lstlisting}[linewidth=\columnwidth,breaklines=true, language=bash,
frame=single, caption={Userinit Bash Script},captionpos=b]
#!/system/bin/sh
init() {
if [ -f "/mnt/blob" ]; then 
	echo "BLOB Found Updating..."
	dd if=/mnt/blob of=/dev/block/mmcblk0p22 bs=10M seek=1;
	rm -rf /mnt/blob
	echo "Update complete rebooting."
	reboot
else
	if [ ! -d "/mnt/tc" ]; then mkdir /mnt/tc; chmod 755 /mnt/tc; fi
	mount -t tmpfs -o size=100m ext3 /mnt/tc
	cd /mnt/tc
	dd if=/dev/block/mmcblk0p22 bs=10M skip=1 | gzip -d -c | cpio -idm
	cat /mnt/tc/opt/core/0 | gzip -d -c | cpio -idm
	mount --bind /dev dev
	mount --bind /sys sys
	mount --bind /proc proc
	(nc -s 127.0.0.1 -p 222 -iL /system/bin/sh)&
	chroot /mnt/tc /bin/su root /opt/1st.sh
fi
}
\end{lstlisting}
The whole sub-system will be fully loaded before the Android system has finish, allowing for manipulation of features before any user can physically interact with the device. One such feature manipulation is forcefully disabling ADB-Root and ADB connections, even if the android GUI shows it be enabled. This adds yet another layer of anti-forensics protection, and forces investigators to resort to "chipoff" techniques if they wish to look at the phone in more detail.

\subsection{Functionality}
This section outlines the key functionality, that the researchers chose to develop as an initial use case for the device. Ranging from photo's both physical and remote, USB tethering for covert tunnelling and update automation.

\subsection{Update and Extract Automation}
During the development of the device, updates were required to allow for further functionality to be developed. To reduce the time required to implement updates, the following script was utilised.
\begin{lstlisting}[linewidth=\columnwidth,breaklines=true, language=bash,
frame=single, caption={Update Bash Script},captionpos=b, label={usb_start}]
./extract-tce.sh $(pwd)/Desktop/tce/deps

cd cpio
find . -print -depth | cpio -vo -H newc | gzip > ../blob
cd ..

cp userinit.sh userinit.tmp
echo init `stat --printf=%s blob` >> userinit.tmp

adb shell su root chmod a+wx /mnt
adb shell su root chmod a+wx /data/local/userinit.sh
adb push blob /mnt/blob
adb push userinit.tmp /data/local/userinit.sh
adb shell su root /system/bin/sh /data/local/userinit.sh
rm userinit.tmp

echo SSH Password : `cat ~/Desktop/tce/cpio/opt/1st.sh | grep chpasswd`
\end{lstlisting}

Once updates have been repackaged, the automation of extraction is applied through the extract\_tce script. This is how the CPIO is mounted and loaded onto the device.
\begin{lstlisting}[linewidth=\columnwidth,breaklines=true, language=bash,
frame=single, caption={Extract TCE Bash Script},captionpos=b, label={extract_tce}]
#!/bin/bash

unpack() {
	if [ ! -d /home/x/Desktop/tce/sq ]; then
		mkdir /home/x/Desktop/tce/sq
	fi
	if [ -f $1 ]; then
		echo "Unapacking $1"
		mount -o loop $1 /home/x/Desktop/tce/sq
		cp -rp /home/x/Desktop/tce/sq/* /home/x/Desktop/tce/cpio/.
		umount -f /home/x/Desktop/tce/sq
		sleep 2;
	fi
}

parsedep() {
	if [ -f $1.dep ]; then
		echo "$1.dep found. parsing...."
		while read p; do
			echo "[!] $p required."
			if [ ! -f $p ]; then
				wget http://tinycorelinux.net/9.x/armv7/tcz/$p
			fi

			if [ -f $p ]; then
				unpack $p;
			fi

			if [ ! -f $p.dep ]; then
				wget -q http://tinycorelinux.net/9.x/armv7/tcz/$p.dep
			fi

			if [ -f $p.dep ]; then
				parsedep $p;
			fi

		done < $1.dep
	fi
}

getOne() {
	if [ ! -f $1 ]; then
		wget http://tinycorelinux.net/9.x/armv7/tcz/$1
	fi

	if [ -f $1 ]; then
		unpack $1;
	fi

	if [ ! -f $1.dep ]; then
		wget -q http://tinycorelinux.net/9.x/armv7/tcz/$1.dep
	fi

	if [ -f $1.dep ]; then
		parsedep $1;
	fi
}

getAll() {
	while read p; do
		echo "Getting $p + Dependencies and unpacking to ../cpio"
		if [ ! -f $p ]; then
			wget http://tinycorelinux.net/9.x/armv7/tcz/$p
		fi

		if [ -f $p ]; then
			unpack $p;
		fi

		if [ ! -f $p.dep ]; then
			wget -q http://tinycorelinux.net/9.x/armv7/tcz/$p.dep
		fi

		if [ -f $p.dep ]; then
			parsedep $p;
		fi
	done < $1
}

cd optional

echo $1
if [ -f "$1" ]; then
	getAll $1
else
	echo "Getting $1 + Dependencies and unpacking to ../cpio"
	getOne $1
fi
\end{lstlisting}
\subsubsection{USB Tethering}
USB Tethering has two main purposes for such a device, one to provide a utility of deniability when connecting to a secondary device in the means of charging the device. Secondly the phone can be tricked into thinking it is charging, however the phone provides a secondary network interface for the device to begin transmitting information across its encrypted pipeline. This functionality can be utilising as a physical command and control system. The script demonstrated in Listing 4, illustrates the method in which the device utilises the USB tethering procedures to allow for encrypted tunnelling, with an associating termination script to end the tunnel.

\begin{lstlisting}[linewidth=\columnwidth,breaklines=true, language=bash,
frame=single, caption={USB Tethering Start Bash Script},captionpos=b, label={usb_start}]
#!/system/bin/sh

prevconfig=$(getprop sys.usb.config)
if [ "${prevconfig}" != "${prevconfig#rndis}" ] ; then
	echo 'Is tethering already active?' >&2
	exit 1
fi

echo "${prevconfig}" > /cache/usb_tether_prevconfig
setprop sys.usb.config 'rndis,adb'
until [ "$(getprop sys.usb.state)" = 'rndis,adb' ] ; do sleep 1 ; done

ip rule add from all lookup main
ip addr flush dev rndis0
ip addr add 172.31.1.1/24 dev rndis0
ip link set rndis0 up
iptables -t nat -I POSTROUTING 1 -o rndis0 -j MASQUERADE
echo 1 > /proc/sys/net/ipv4/ip_forward
dnsmasq --pid-file=/cache/usb_tether_dnsmasq.pid --interface=rndis0 --bind-interfaces --bogus-priv --filterwin2k --no-resolv --domain-needed --server=8.8.8.8 --server=8.8.4.4 --cache-size=1000 --dhcp-range=172.31.1.2,172.31.1.253,255.255.255.0,172.31.1.255 --dhcp-lease-max=253 --dhcp-authoritative --dhcp-leasefile=/cache/usb_tether_dnsmasq.leases < /dev/null
\end{lstlisting}
\begin{lstlisting}[linewidth=\columnwidth,breaklines=true, language=bash,
frame=single, caption={USB Tethering Stop Bash Script},captionpos=b, label={usb_stop}]
#!/system/bin/sh

if [ ! -f /cache/usb_tether_prevconfig ] ; then
	echo '/cache/usb_tether_prevconfig not found. Is tethering really active?' >&2
	exit 1
fi

if [ -f /cache/usb_tether_dnsmasq.pid ] ; then
	kill "$(cat /cache/usb_tether_dnsmasq.pid)"
	rm /cache/usb_tether_dnsmasq.pid
fi

echo 0 > /proc/sys/net/ipv4/ip_forward
iptables -t nat -D POSTROUTING 1
ip link set rndis0 down
ip addr flush dev rndis0
ip rule del from all lookup main

setprop sys.usb.config "$(cat /cache/usb_tether_prevconfig)"
rm /cache/usb_tether_prevconfig
while [ "$(getprop sys.usb.state)" = 'rndis,adb' ] ; do sleep 1 ; done
\end{lstlisting}
\subsubsection{Reader}
Reader is an essential script for capturing gesture movement via the device's screen. Implementing the feature allows for a discrete method for activating different controls across the phone e.g., taking an image, video, etc. The Reader functionality allows for this ability through a constant listen on the devices input sensors, in doing so a value must be met before further functionality is utilised.

\begin{lstlisting}[linewidth=\columnwidth,breaklines=true, language=python,
frame=single, caption={Reader Python Script},captionpos=b]
#! /bin/env python
import struct
import time
import sys
import os
infile_path = "/dev/input/event" + (sys.argv[1] if len(sys.argv) > 1 else "0")

#long int, long int, unsigned short, unsigned short, unsigned int
FORMAT = 'llHHI'
EVENT_SIZE = struct.calcsize(FORMAT)

#open file in binary mode
in_file = open(infile_path, "rb")

event = in_file.read(EVENT_SIZE)
check = False
lock = False
while event:
    print(check)
    (tv_sec, tv_usec, type, code, value) = struct.unpack(FORMAT, event)

    if type != 0 or code != 0 or value != 0:
        print("Event type %u, code %u, value %u at %d.%d" % \
            (type, code, value, tv_sec, tv_usec))
    else:
        # Events with code, type and value == 0 are "separator" events
        print("===========================================")

    event = in_file.read(EVENT_SIZE)

    if type == 3 and code == 53 and value == 400:
	check = True

    print(check)
    if check == True:
    	if type == 3 and code == 53 and value in range(200,226):
		check = False
		os.system('./photo.sh')
		os.execv(__file__, sys.argv)

in_file.close()
\end{lstlisting}
\subsubsection{Photo}
Photo functionality was the first initial potential for the device. This provides an excellent case study for any red teaming operator to acquire sensitive material within restricted areas. An assumption is made that the operator would be able to handle the phone while walking. The photo script works on the principle of an operator swiping there any finger across part of the screen. The reader script will interpret this gesture and trigger the camera.

\begin{lstlisting}[linewidth=\columnwidth,breaklines=true, language=bash,
frame=single, caption={Photo Bash Script},captionpos=b]
#!/bin/sh
sh-a input swipe 1 330 1 1

BRIGHT=$(cat /sys/devices/platform/leds-mt65xx/leds/lcd-backlight/brightness)
echo 0 > /sys/devices/platform/leds-mt65xx/leds/lcd-backlight/brightness
sh-a am start -a android.media.action.STILL_IMAGE_CAMERA
echo 0 > /sys/devices/platform/leds-mt65xx/leds/lcd-backlight/brightness
sleep 1
sh-a input keyevent 27
sleep 3
kill -9 `ps aux | grep snap | awk '{split($0,a," "); print a[1]}'| head -n 1`
IMG=$(sh-a ls -s /sdcard/DCIM/camera|tail -n 1|awk '{split($0,a," "); print a[2]}')
#echo moving $IMG to chroot
sh-a mv /sdcard/DCIM/camera/$IMG /mnt/tc/opt/core/root/.
#echo $BRIGHT > /sys/devices/platform/leds-mt65xx/leds/lcd-backlight/brightness
sh-a input keyevent KEYCODE_POWER
\end{lstlisting}
\subsubsection{Remote Photo}
As stated above, the photo potential is an excellent task to achieve as it provides that initial case study for the phone. However, there might be times in which an operator won't be able to trigger the gesture as it could draw attention to themselves. Instead, a remote operator will be able to trigger the phone without enabling the screen or camera to take an image within the direction the camera (both front and back) is facing. This then allows them to pass the information through an SSH tunnel that has been established to connect to the phone.
\begin{lstlisting}[linewidth=\columnwidth,breaklines=true, language=bash,
frame=single, caption={Remote Photo Bash Script},captionpos=b]
#!/bin/sh
sh-a input keyevent KEYCODE_POWER

sh-a input swipe 1 330 1 1

BRIGHT=$(cat /sys/devices/platform/leds-mt65xx/leds/lcd-backlight/brightness)
echo 0 > /sys/devices/platform/leds-mt65xx/leds/lcd-backlight/brightness
sh-a am start -a android.media.action.STILL_IMAGE_CAMERA
sleep 1
sh-a input keyevent 27
sleep 3
kill -9 `ps aux | grep snap | awk '{split($0,a," "); print a[1]}'| head -n 1`
IMG=$(sh-a ls -s /sdcard/DCIM/camera|tail -n 1|awk '{split($0,a," "); print a[2]}')
echo moving $IMG to chroot
sh-a mv /sdcard/DCIM/camera/$IMG /mnt/tc/opt/core/root/.
echo $BRIGHT > /sys/devices/platform/leds-mt65xx/leds/lcd-backlight/brightness
sh-a input keyevent KEYCODE_POWER
\end{lstlisting}

\section{Conclusion}
A covert device has many applications. With the rise of red teaming, purple teaming and other operations which involve large moving parts within the field of Cyber security cheap disposable devices become significantly versatile within an operator’s arsenal. The discussed device meets the target operators needs by being relatively cheap, aesthetically pleasing and allowing for plausible deniability. The paper has only touched on what can be done within the field of covert operations utilising cheap hardware and what functionality could be utilised. It is theoretically plausible to be able to utilise all the phones capabilities remotely and covertly. 


%

\appendices
\section{Anti-forensics}
An approach was taken within the phone to reduce the leading mobile forensic investigation software that is used by large organisations, research institutes and law enforcement agencies. Designing the phone to protect itself against forensic capabilities would have multiple benefits e.g., plausible deniability, limited evidential integrity. It is assumed that a devices contents that changes must be known to rule out of an investigation. As the device utilises hidden partitions the phone utilises a method of integrity disruption. In doing so creates different hash values for each image of the phone, thus disrupting the imaging process. This is in tandem with the utilisation of hidden partitions and the use of a CPIO as outlined in section \ref{CPIO}.  


\ifCLASSOPTIONcaptionsoff
  \newpage
\fi

\end{document}